\documentclass[a4paper]{jpconf}
\usepackage{graphicx}
\usepackage{amsmath}

\begin{document}
\title{Vacuum properties of high quality value tuning fork in high magnetic field up to 8 Tesla and at mK temperatures}

\author{M \v Clove\v cko, M Kupka, P Skyba and F Vavrek}

\address{Centre of Low Temperature Physics, Institute of Experimental Physics SAS and P J \v Saf\'arik University Ko\v sice, Watsonova 47, Ko\v sice, Slovakia}

\ead{clovecko@saske.sk}

\begin{abstract}
Tuning forks are very popular experimental tools widely  applied in low  and ultra low
temperature physics as mechanical resonators and cantilevers in the study of quantum liquids,
STM and AFM techniques, etc. As an added benefit, these forks being cooled, have very high
Q-value, typically $10^6$ and their properties seems to be magnetic field independent. We
present  preliminary vacuum measurements of a commercial tuning fork oscillating at frequency
32~kHz conducted in magnetic fields up to 8~T and at temperature $\sim 10$~mK.  We found an
additional weak damping of the tuning fork motion depending on magnetic field magnitude and we
discuss  physical nature of the observed phenomena.
\end{abstract}

\section{Introduction}

Quartz tuning forks are versatile mechanical resonators with really high Q-values responsible
for their superior sensitivity. No wonder that they are applied in all sub-fields of low
temperature physics in the study of quantum liquids and solids, quantum turbulence, etc.~\cite{Clubb04, Blaauwgeers2007, Blazkova07, Blazkova08, Pentti08, Bradley09, Schmoranzer11, ahlstrom} and even as the important part in scanning
probe techniques like AFM, STM, etc.~\cite{STM, Rych1, Rych2}. In the latter case the fork
measurements are often carried out in vacuum and at low temperatures including strong magnetic
fields applied. Despite its obvious importance, the influence of strong magnetic fields on the
vacuum resonant properties of such tuning fork at mK temperature range (to our knowledge) has
not been studied yet. Therefore, the main aim of our work was to investigate the response of
tuning fork under such conditions.

\section{Experimental details}

\begin{figure}
\begin{tabular} {c c}

\hspace{33pt}
\includegraphics[height = 60mm]{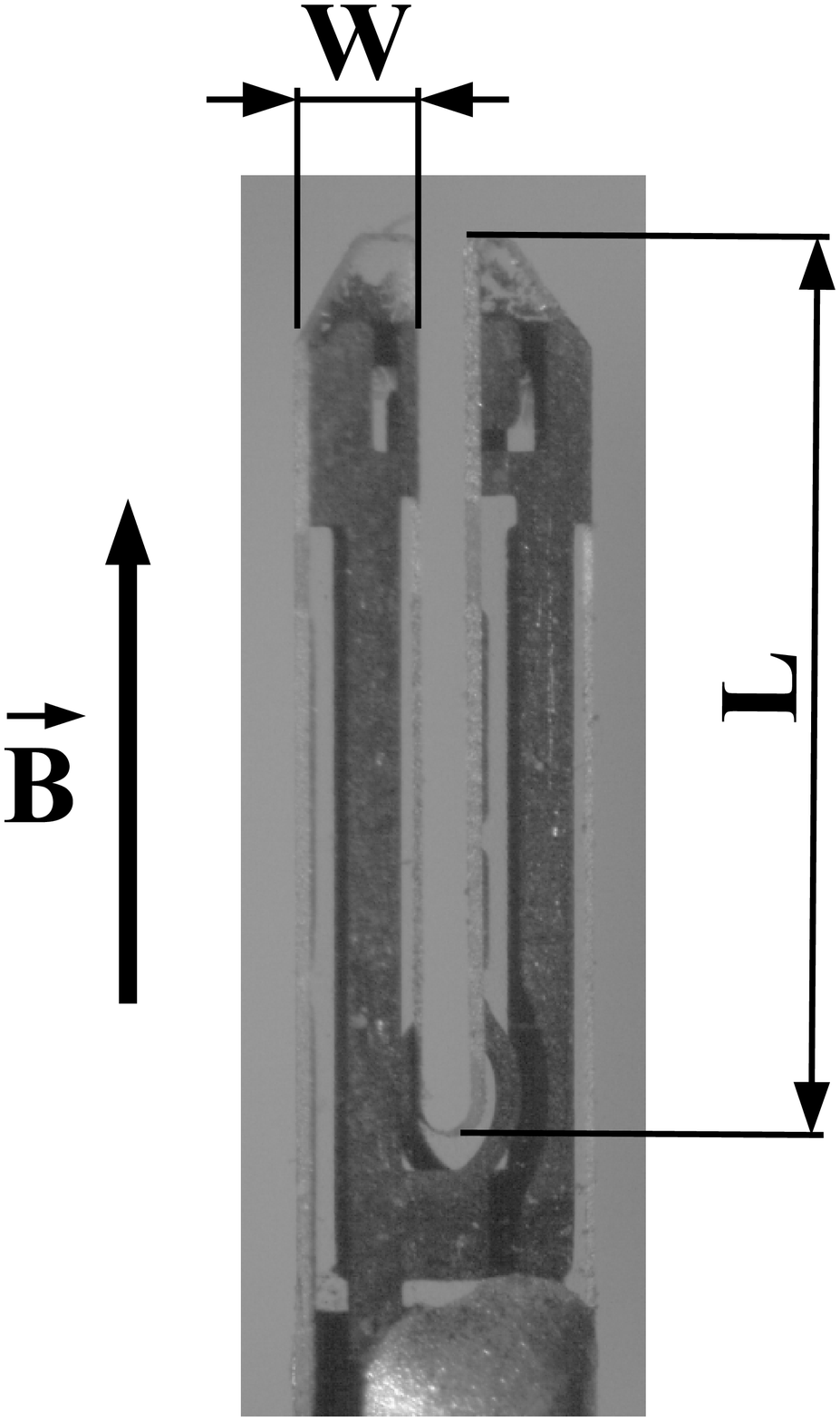}
\hspace{20pt}

&

\includegraphics[width = 0.57\textwidth]{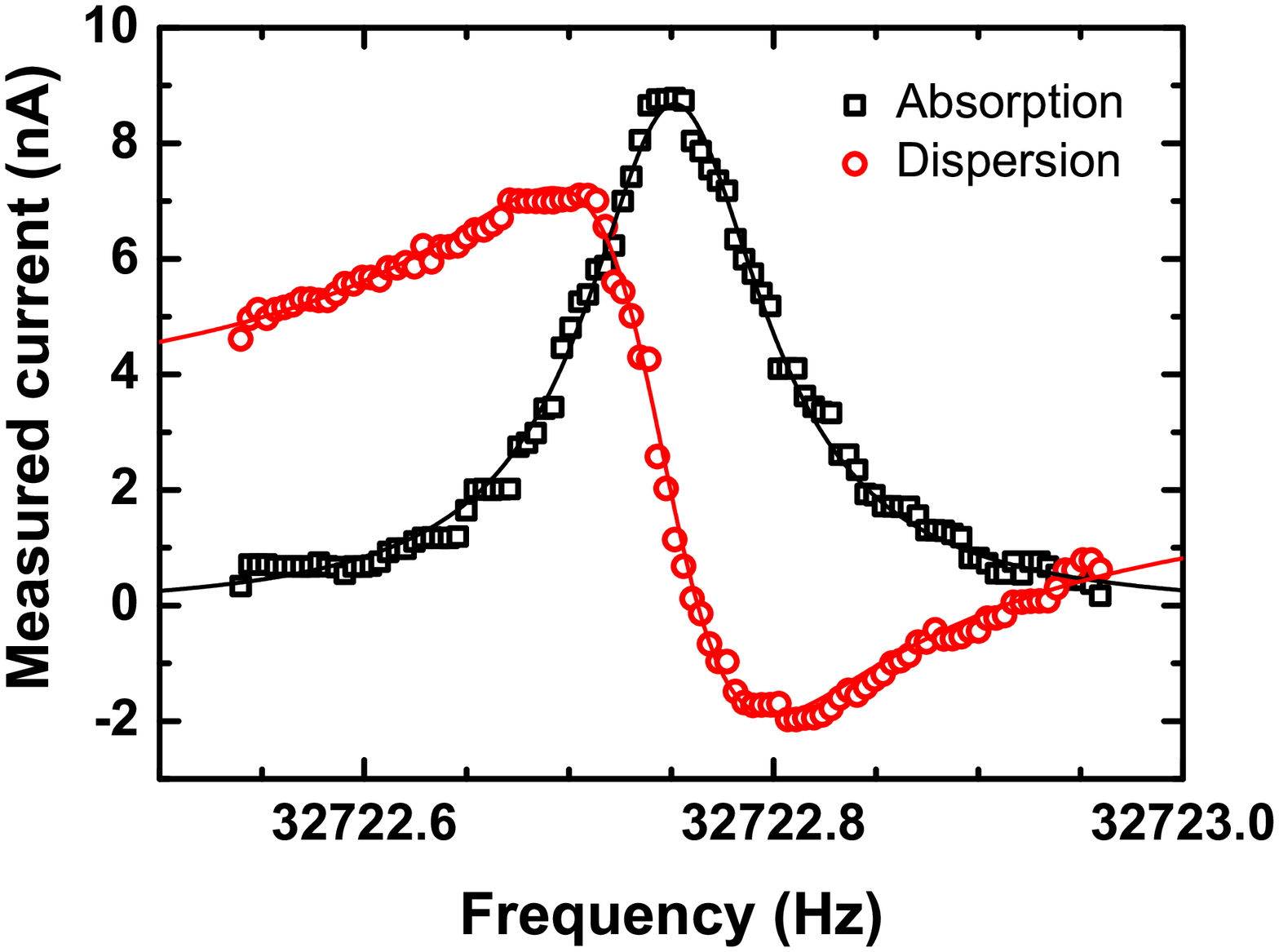}
\hspace{-40pt}
\vspace{-8pt}

\end{tabular}

\caption{\textit{Left:} 32 kHz quartz tuning fork used with the prong width $W = 400$ $\mu$m and
length $L = 3117$ $\mu$m. The wafer thickness $T=230$ $\mu$m. \textit{Right:} A typical
frequency dependence of piezoelectric current (resonant curves) measured. Lines represents fits of experimental data with the resulting quality factor $Q\sim 5\times 10^5$.} \label{fig1}
\end{figure}

To perform our measurements a commercially available quartz tuning fork resonating at $\sim$~32~kHz was used.
In order to be able to measure the tuning fork in strong magnetic fields a~few
modifications have been made. Firstly, the metal can was removed and magnetic leads were
replaced with non-magnetic ones (twisted thin copper wire pair in our case). Once out of the
can, the dimensions of the tuning fork were measured by an optical microscope (see
fig.\ref{fig1}). Finally, the bare tuning fork was fixed on a specially designed copper holder
and this whole setup was mounted on the cold finger of our cryogen-free dilution refrigerator
Oxford Triton~200, which is capable to cool our samples down to 10 mK in the magnetic fields up
to 8 T. The orientation of tuning fork's prongs was parallel to the applied magnetic field.

The response of tuning fork in the form of piezoelectric current on driving force was
measured by verified technique \cite{Blaauwgeers2007, notes}.  The driving force i.e. an
AC-voltage slowly swept in frequency  was provided by the function generator Agilent~33521A. An
additional attenuator attenuated this excitation voltage by 40 dB. When driven with AC voltage
at frequency close to the fork resonant frequency, the quartz crystal starts to oscillate and
the piezoelectric current flows as a direct consequence of periodic changes of the crystal
lattice polarization in time. This resulting piezoelectric current was detected and converted to
voltage by home-made current-to-voltage (I/V) converter with gain 10$^5$ V/A \cite{IVconv}.
The output voltage signal from I/V converter was measured by the phase-sensitive (lock-in)
amplifier SR 830, which splits the measured signal into two phase components: in phase
(absorption) component and quadrature (dispersion) component relative to the reference signal
provided by above mentioned function generator.

The typical resonant curves are shown on the right side of fig.~\ref{fig1}. Experimental data can
be fitted using well known Lorentz relations
\begin{align}
I_\mathrm{abs} & = I_0\frac{(f \Delta f)^2}{(f_0^2-f^2)^2+(f \Delta f)^2} \\
I_\mathrm{dis} & = I_0\frac{f \Delta f (f_0^2-f^2)^2}{(f_0^2-f^2)^2+(f \Delta f)^2} \label{equ1}
\end{align}
to obtain the resonant frequency $f_0$, the frequency linewidth $\Delta f$ and the amplitude of
piezoelectric current~$I_0$. There are still some constant and linear backgrounds in actual
measurements present, which were taken into account as well \cite{notes}. The piezoelectric
current~$I_0$ is directly proportional to the velocity $v$ of prong tip of tuning fork. The
constant of proportionality is the tuning fork constant $\alpha$ and can be determined
experimentally using following expression~\cite{Blaauwgeers2007}:
\begin{equation}
\alpha = \sqrt{\frac{2\, m_\mathrm{vac}\Delta\omega}{R}}.
\label{alpha}
\end{equation}
Here $\Delta\omega = 2\pi\Delta f$ is the angular linewidth, $R$ is the resistance of tuning
fork at resonance, that models the damping of the fork motion and can be obtained from the fit of $I_0$ vs. the amplitude of driving AC
voltage $U_\mathrm{exc}$. Finally, $m_\mathrm{vac} = 0.25 \,\rho \,TWL$ is the effective mass of
one fork's prong (quartz density $\rho = 2659$~kg.m$^{-3}$). The resulting effective mass for
our tuning fork $m_\mathrm{vac} = 1.90645\times 10^{-7}$~kg.

\begin{figure}
\begin{tabular} {c c}

\hspace{-25pt}
\includegraphics[width = 0.57\textwidth]{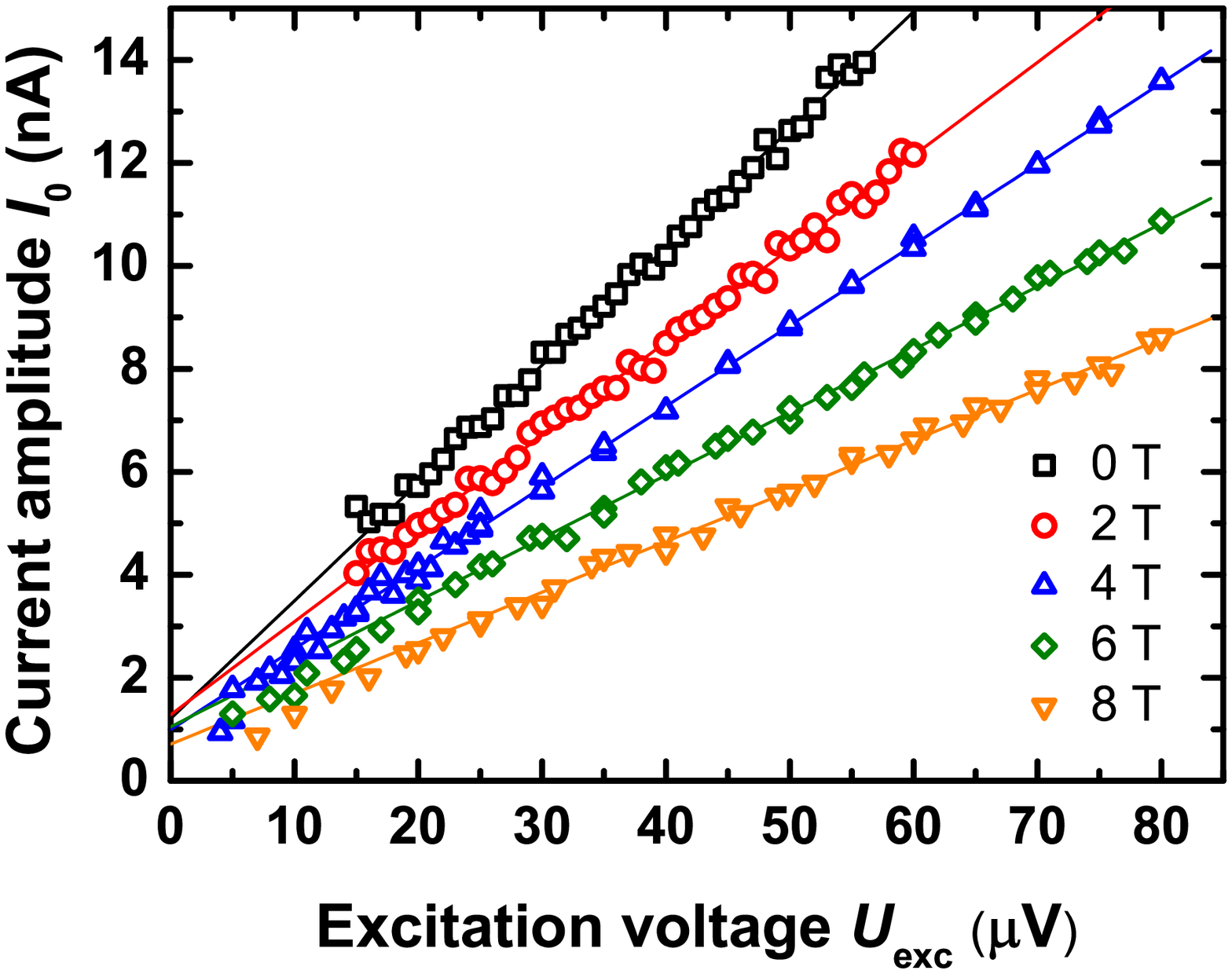}
\hspace{-20pt}

&

\hspace{-20pt}
\includegraphics[width = 0.57\textwidth]{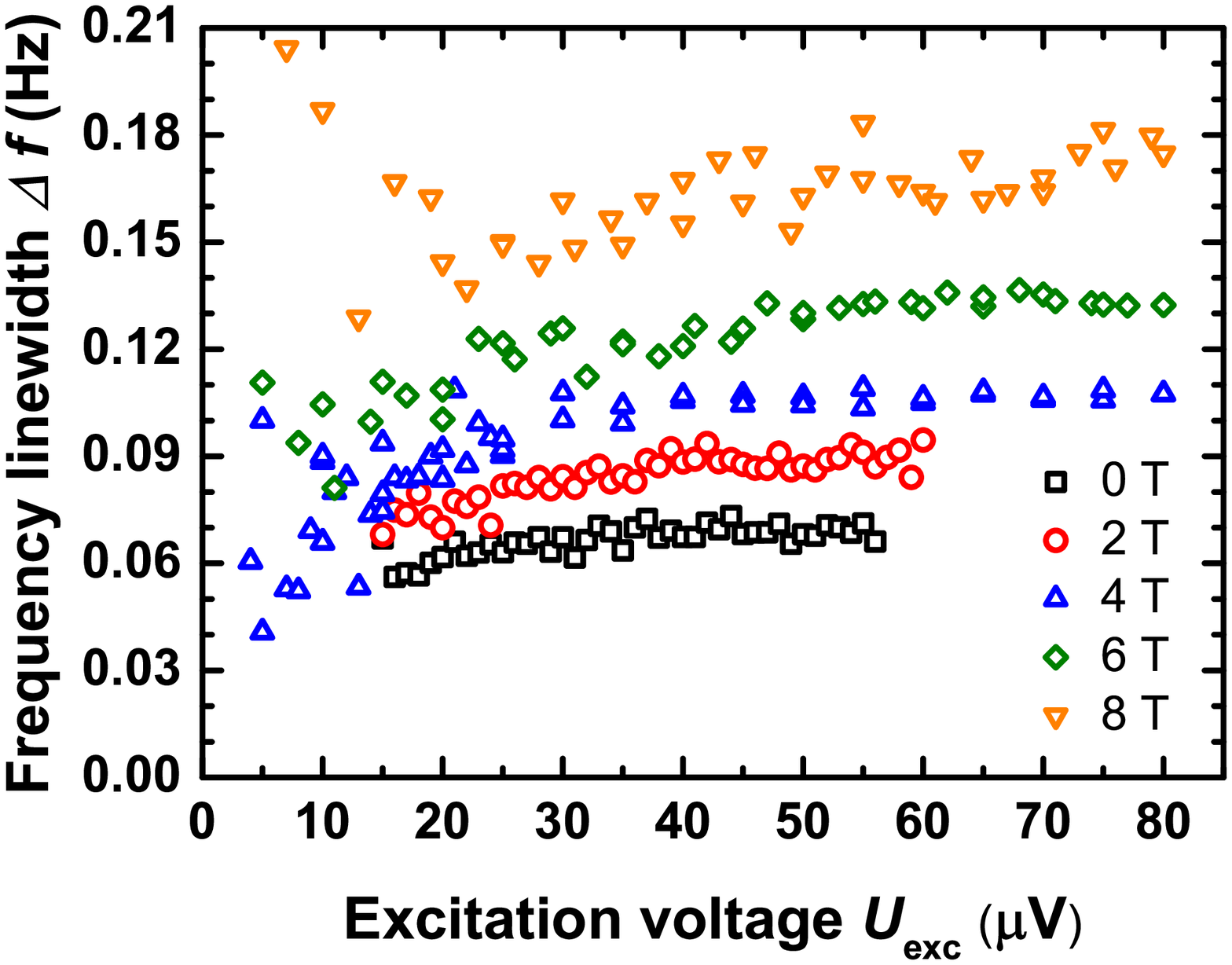}
\hspace{-40pt}

\end{tabular}

\caption{\textit{Left:} Dependencies of the piezoelectric current amplitude $I_0$ on the
excitation voltage~$V_\mathrm{exc}$ and magnetic field applied. \textit{Right:} Frequency
linewidth $\Delta f$ as a function of the excitation voltage $V_\mathrm{exc}$ and magnetic field
applied.} \label{fig2}
\end{figure}

\section{Results and discussion}

All measurements were performed in vacuum at $\sim 10$~mK.  Figure \ref{fig2} shows dependencies
of the piezoelectric current amplitude (left) and frequency linewidth~$\Delta f$ (right) on
excitation voltage amplitude $V_\mathrm{exc}$ as measured at different values of magnetic field.
The dependence of current amplitude~$I_0$ on excitation voltage amplitude $V_\mathrm{exc}$
suggests that the tuning fork resistance~$R$ (i.e. the damping of the fork motion) is rising
with the increase of magnetic field (the corresponding slope - the conductance showed in fig
\ref{fig2} is decreasing). Similarly, the frequency linewidth~$\Delta f$ is increasing as well.
As measurements were carried out in vacuum, there is no influence of an external environment on
the tuning fork motion and the tuning fork resistance in zero magnetic field $R_\mathrm{0}$
reflects only an intrinsic damping processes. This intrinsic process of energy dissipation can
be associated with shear friction as consequence of periodically bending crystal lattice of
quartz.

The linear fits to the experimental data show a non-zero offset and there is a slight deviation
of frequency linewidth $\Delta f$ observed at small excitation voltages~$U_\mathrm{exc} \leq
30$~$\mu$V. Both these phenomena can be attributed to the contribution of the TTL logic (used as
a reference signal for lock-in amplifier) to the excitation voltage $U_\mathrm{exc}$ due to the
presence of capacitive coupling \cite{notes}. Thus, a non-zero piezoelectric current $I_0$ can
also be detected even for zero excitation voltage amplitude $U_\mathrm{exc}$. However, the
corresponding signal measured by lock-in amplifier is continuously shifted from 0$^\circ$ to 90$^\circ$ in phase as $U_\mathrm{exc} \rightarrow 0$~V. This also explains deviation of the measured linewidths~$\Delta f$ from constant value for low excitation voltage amplitudes.

\begin{figure}
\vspace{-15pt}
\includegraphics[width = 0.57\textwidth]{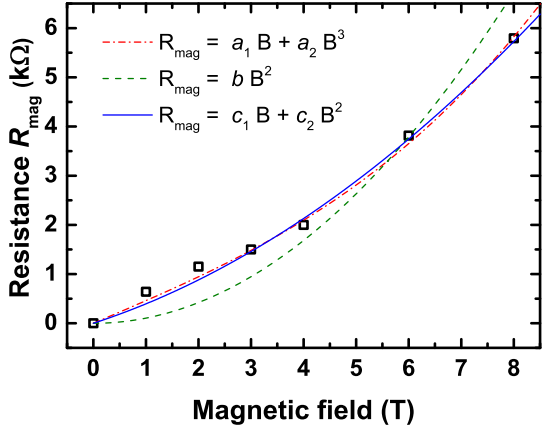}
\begin{minipage}[b]{14pc}
\caption{\label{fig3} Magnetic contribution to the tuning fork resistance $R_\mathrm{mag}$
plotted against magnetic field applied. Presented fits correspond to dipole twist ($\chain$), pure eddy currents ($\broken$) and combination of both ($\full$).}
\end{minipage}
\vspace{-15pt}
\end{figure}

The increase of the fork resistance $R$ with the rising magnitude of magnetic field indicates
the presence of of additional damping mechanism acting on tuning fork motion. Assuming that
intrinsic damping process in zero field ($R_\mathrm{0}$) i.e. the shear friction is magnetic
field independent, a magnetic contribution to the tuning fork resistance $R_\mathrm{mag}$ can be
simply estimated as $R_\mathrm{mag} = R_\mathrm{B} - R_\mathrm{0}$. The fig. \ref{fig3}
illustrates the resulting dependence of $R_\mathrm{mag}$ on applied magnetic field. What could
be an origin of the additional, the field depended damping in quartz tuning fork?

The forced motion of one prong of the tuning fork pointing in the direction of z axis in zero magnetic field is described by the
Euler-Bernoulli equation
\begin{equation}
\rho A \frac{\partial^2 x}{\partial t^2} + 2 \gamma \rho A \frac{\partial x}{\partial t} + E_m
J_y \frac{\partial ^4 x}{\partial z^4} = F(z) \cos (\omega t), \label{equ2}
\end{equation}
where $A$ is the arm cross-section, $\rho$ is the density, $E_m$ is the Young's modulus of the
material, $J_y$ is the moment of inertia of the prong cross-section and $\gamma$ characterizes
the damping process. The external force $F(z)$ is provided by the AC-voltage applied on
electrodes of the tuning fork and resulting harmonic electric field $\mathbf E$ creates  the
time dependent charge polarization ${\mathbf p}(t)$ of the crystal lattice of the tuning fork.
This time dependent charge polarization $\dot{\mathbf p}(t)$ i.e.~the piezoelectric current
carries information about fork motion.

Once the magnetic field is applied on tuning fork, in general, there are two effects acting
simultaneously on dynamics of the tuning fork motion. The first, in our geometry - magnetic
field applied in parallel with prongs of the tuning fork,  magnetic field affects the excitation
force provided by electric field $\mathbf E$ by adding a force acting perpendicularly to the
former one ($q {\mathbf E} + \dot{\mathbf p} \times {\mathbf B}$). This additional Lorentz force
has a tendency to twist the oscillating dipole moments inside quartz crystal from the direction
of electric field $\mathbf E$. This effect effectively reduces the magnitude of the
piezoelectric current and thus increases the fork's resistance. The ratio $(\dot{\mathbf p}
\times {\mathbf B})/q {\mathbf E}$ defines a tangent of an angle ($\theta$), by which the
oscillating dipoles are twisted. Then, using this simplest physical picture and assuming the
smallness of $\theta$, the magnetic contribution to the tuning fork resistance can be expressed
in form $R_\mathrm{mag}= a_1 B + a_2 B^3$, where $a_1$ and $a_2$ are constants. The second effect is a
damping associated with generation of the eddy currents in fork's metallic electrodes of during
its oscillating motion. An EDAX analysis of the electrodes showed that electrodes consist of Cr,
Ag and Sn. The damping caused by eddy currents is proportional to $b\, B^2$.

The lines in fig. \ref{fig3} show the fits to the experimental data considering each mechanism (dipole twist and eddy currents) separately and acting together (without $B^3$ term). As follows from data fits the contribution due to the eddy currents does not fit experimental data properly, so additional contribution originating in dipole twist needs to be considered as well. Moreover, the dipole twist mechanism is capable to describe our experimental data by itself.
Both above mentioned mechanisms should depend on relative orientation of the tuning fork and
magnetic field, which could help to discriminate these two effects. However, the most important fact
which comes out from data analysis is that in our configuration (tuning fork's prongs orientated in parallel with magnetic field), the magnitude of the $\alpha$ constant is almost independent on magnetic field within error of $\sim$ 5\% (see Table 1.).

\begin{table}
\caption{Values (and their errors) determined from experiment for frequency linewidth $\Delta
f$, tuning fork resistance $R$ and calculated values (errors) of fork constant $\alpha$
according to relation~\eqref{alpha} for different magnetic fields applied.}
\begin{center}
\begin{tabular}{*{7}{c}}
\br
$B$ & $\Delta f$    & $\Delta f_{err}$          & $R$           & $R_{err}$             & $\alpha$          & $\alpha_{err}$            \\
T       & Hz                    & $\pm$ Hz                          & $\Omega$  & $\pm$ $\Omega$    & A.s.m$^{-1}$  & $\pm$ A.s.m$^{-1}$    \\
\mr
$0$ &   $0.06894$   &   $0.00225$   &   $4370.59876$    &   $57.15386$  &   $6.14709\times 10^{-6}$ &   $1.08088\times 10^{-7}$ \\
$1$ &   $0.07905$   &   $0.00237$   &   $5011.08829$    &   $76.42912$  &   $6.14755\times 10^{-6}$ &   $1.03427\times 10^{-7}$ \\
$2$ &   $0.08885$   &   $0.00271$   &   $5523.22733$    &   $84.74155$  &   $6.20787\times 10^{-6}$ &   $1.05905\times 10^{-7}$ \\
$3$ &   $0.09695$   &   $0.00261$   &   $5871.39784$    &   $39.92771$  &   $6.28943\times 10^{-6}$ &   $8.72780\times 10^{-8}$ \\
$4$ &   $0.10649$   &   $0.00151$   &   $6369.90292$    &   $40.71070$  &   $6.32847\times 10^{-6}$ &   $4.91689\times 10^{-8}$ \\
$6$ & $0.13341$ &   $0.00165$   &   $8185.47428$    &   $85.27016$  &   $6.24880\times 10^{-6}$ &   $5.04657\times 10^{-8}$ \\
$8$ &   $0.17520$   &   $0.00448$   &   $10164.13935$   &   $94.14669$  &   $6.42605\times 10^{-6}$ &   $8.73431\times 10^{-8}$ \\
\br
\end{tabular}
\end{center}
\end{table}

\section{Conclusions}
We have presented that the fork constant $\alpha$ of our $32$~kHz quartz tuning fork is almost
independent on the magnetic field applied. This result can be of large importance for
measurements performed in high magnetic fields, e.g. in various scanning probe techniques
utilizing quartz tuning forks as probes. The origin of additional damping of tuning fork motion
due to the presence of magnetic field is still an open question. To elucidate this problem more experiments with different relative orientation of tuning fork's prongs with respect to magnetic field are needed.

\ack We would like to thank to V. Komanick\'y for performing the spectral analysis of the
tuning fork. CLTP is operated as the CFNT MVEP of the Slovak Academy of
Sciences and P. J. \v{S}af\'arik University. We acknowledge support provided by grants:
APVV-0515-10, VEGA 2/0128/12, CEX-Extrem ITMS 26220120005 (SF of EU) and the Microkelvin, project of
7th FP of EU. Support provided by the U.S. Steel Ko\v{s}ice s.r.o. is also very appreciated.

\end{document}